\documentclass[letter]{aa}  
\usepackage[colorlinks=true,allcolors=blue]{hyperref}
\usepackage{graphicx}
\usepackage{txfonts}
\newcommand{\kms}{km\,s$^{-1}$}

\newcommand{\vsini}{$v_{\rm e} \sin i$}
\newcommand{\teff}{$T_{\rm eff}$}
\newcommand{\logg}{$\log{g}$}
\newcommand{\bz}{$\langle B_{\rm z} \rangle$}
\newcommand{\bs}{$\langle B \rangle$}
\newcommand{\hd}{HD\,57372}

\defcitealias{hubrig:2024}{H24}

\newcommand{\figps}[3]{\resizebox{#1}{!}{\rotatebox{#2}{\includegraphics{#3}}}}
\newcommand{\cla}[1]{{#1}}

\begin{document}

\title{Extremely asymmetric bipolar magnetic field \\ of the Bp star HD\,57372%
\thanks{Based on data obtained from the ESO Science Archive Facility.}
}
%\subtitle{}

\titlerunning{Asymmetric magnetic field of HD\,57372}

\author{Oleg Kochukhov}

\institute{
Department of Physics and Astronomy, Uppsala University, Box 516, 75120 Uppsala, Sweden\\
\email{oleg.kochukhov@physics.uu.se}
}

\date{Received 6 March 2025 / Accepted 8 April 2025}

\abstract{
Fossil magnetic fields of early-type stars are typically characterised by symmetric or slightly distorted oblique dipolar surface geometries. Contrary to this trend, the late-B magnetic chemically peculiar star \hd\ exhibits an unusually large rotational variation of its mean magnetic field modulus, suggesting a highly atypical field configuration. In this study, we present a Zeeman Doppler imaging analysis of \hd, revealing an exceptionally asymmetric bipolar magnetic topology, rarely observed in early-type stars. According to our magnetic field maps, reconstructed from the intensity and circular polarisation profiles of Fe, Cr, and Ti lines, approximately 66 per cent of the stellar surface is covered by a diffuse outward-directed radial field, with local field strengths reaching 11.6 kG, while the remaining 34 per cent hosts a highly concentrated inward-directed field with a strong horizontal component and a peak strength of 17.8 kG. These unusual surface magnetic field characteristics make \hd\ a notable object for testing fossil-field theories and interpreting phase-resolved spectropolarimetric observations of early-type stars.
}


\maketitle

\section{Introduction}

Magnetic fields play a fundamental role in the physics of stars, influencing their structure, evolution, and interaction with their surroundings. In about 7--10 per cent of early-type stars -- those of spectral types O, B, and A -- the presence of large-scale surface magnetic fields has been firmly established through spectropolarimetric observations \citep{donati:2009,morel:2015,wade:2016,grunhut:2017}. Unlike the turbulent, convective dynamos responsible for magnetic activity in late-type stars, early-type stars lack significant outer convection zones, and their magnetic fields are believed to be fossil remnants of primordial magnetic flux trapped during star formation or generated by early binary mergers \citep{braithwaite:2004,schneider:2019}. These fields can strongly affect stellar winds, rotation, and angular momentum evolution, leading to complex magnetospheric structures \citep{petit:2013,keszthelyi:2023}.

The majority of magnetic early-type stars exhibit relatively simple, large-scale field geometries, most commonly approximated by a tilted dipole, with strengths ranging from a few hundred G to tens of kG \citep{auriere:2007,shultz:2018a,sikora:2019a}. These fields are typically detected through Zeeman splitting and circular polarisation signatures in spectral lines, observed using high- or medium-resolution spectropolarimetric techniques \citep[e.g.][]{bagnulo:2015,mathys:2017,chojnowski:2019}. Rotational modulation of spectral line profiles and polarisation signatures enables the determination of detailed surface field geometries. This can be achieved either by fitting integral magnetic observables, such as the longitudinal magnetic field \bz\ and the field modulus \bs, using simple parameterised field topology models \citep{landstreet:2000,bagnulo:2002}, or by reconstructing detailed maps of the surface field and associated chemical star spots using the full Stokes parameter profile information with the Zeeman Doppler Imaging (ZDI) inversion technique \citep{kochukhov:2016}.

The application of these different magnetic field modelling methodologies reveals a complex and nuanced picture. It is well known that the sinusoidal rotational \bz\ curves of early-type stars can be successfully described by simple dipolar geometries, leading to the conclusion that such field configurations dominate in most known cases \citep[e.g.][]{auriere:2007,sikora:2019a}. However, interpreting the rotational variation of multiple magnetic observables or reproducing non-sinusoidal \bz\ curves necessitates the introduction of more complex magnetic topologies, prompting claims of ubiquitous contributions of quadrupolar field components \citep[e.g.][]{landstreet:1990,landstreet:2000,bagnulo:2002}. 

ZDI analyses, in particular, have demonstrated common departures from simple dipolar field topologies, revealing that most stars exhibit distortions, asymmetries, and smaller-scale structures superimposed on the basic dipolar configuration \citep{kochukhov:2004d,kochukhov:2019,kochukhov:2022,kochukhov:2023a,kochukhov:2010,silvester:2014}. At the same time, ZDI studies have failed to confirm the presence of quadrupole-dominated fields in early-type stars, instead finding asymmetric dipolar structures even in stars with distinctly non-sinusoidal \bz\ curves \citep{silvester:2015,kochukhov:2017a,rusomarov:2018,semenko:2024}. \cla{Two early-B magnetic stars, HD\,37776 \citep{kochukhov:2011a} and $\tau$~Sco \citep{donati:2006b,kochukhov:2016a}, stand out as notable exceptions, exhibiting highly complex surface magnetic fields dominated by harmonic components with $\ell\ge3$.} Understanding this diversity of field geometries is crucial for refining theories of stellar magnetism and its role in stellar evolution.

In this discussion, the late-B star \hd\ emerges as a particularly intriguing object for studying deviations from a simple dipolar magnetic field geometry. This relatively faint chemically peculiar star, classified as B8p Si by \citet{renson:2009}, has frequently been the subject of photometric variability studies \citep{bernhard:2015,netopil:2017,jagelka:2019}, which documented its high-amplitude, double-wave variation over a 7.889-day rotational period. \citet[][hereafter H24]{hubrig:2024} conducted the first spectroscopic and spectropolarimetric analysis of \hd, reporting the detection of Zeeman-split lines in medium-resolution near-infrared APOGEE spectra as well as in high-resolution optical HARPS observations. These observations revealed a magnetic field modulus reaching up to 15.6~kG, with strong variations over the rotational cycle. \citetalias{hubrig:2024} also measured the longitudinal magnetic field changing between $-6$ and 1.7~kG.

Notably, \hd\ exhibits an exceptionally large modulation of its mean field modulus with a staggering 10~kG difference between extrema in its \bs\ phase curve, as determined by \citetalias{hubrig:2024} from Gaussian fits to the resolved components of several Zeeman-split spectral lines. This variation, corresponding to \bs$_{\rm max}/$\bs$_{\rm min}$\,$\approx$\,3, significantly exceeds the expected factor of $\leq$\,1.25 for a centred dipolar field \citep{kochukhov:2024}. Consequently, \citetalias{hubrig:2024} speculated that \hd\ possesses a very unusual magnetic field geometry. However, no quantitative magnetic field model was proposed in their study, and an attempt to derive dipolar field parameters using conventional formulas yielded no useful results.

The goal of the present work is to construct a detailed model of the magnetic field geometry of \hd\ based on high-resolution spectropolarimetric observations obtained by \citetalias{hubrig:2024}. Through this analysis, we aim to characterise the star's surface magnetic field, assess the validity of claims regarding its complexity, and place it within the broader context of field geometry studies of early-type stars.

\section{\cla{Archival} spectropolarimetric observations}

Eight observations of \hd\ were obtained with the HARPS spectrograph \citep{mayor:2003} at the ESO 3.6-m telescope between 23 April and 1 May 2022. The instrument was used in combination with the circular polarimetric unit, HARPSpol \citep{piskunov:2011}, enabling the recording of Stokes $I$ and $V$ spectra with a resolving power of 110\,000 and a wavelength coverage from 379 to 691~nm, except for a gap in the 526--534~nm region. A detailed log of these observations can be found in \citetalias{hubrig:2024}.

We retrieved the raw HARPSpol spectra of \hd, along with the associated calibration exposures, from the ESO archive\footnote{\url{http://archive.eso.org/eso/eso_archive_main.html}}. These data were processed using the \textsc{REDUCE} echelle spectral reduction package \citep{piskunov:2002}, following the procedures outlined in our previous publications \citep[e.g.][]{rusomarov:2013,rusomarov:2018}.

\section{Stellar parameters}
\label{sec:params}

\citetalias{hubrig:2024} adopted \teff\,=\,13\,500~K and \logg\,=\,4.0 in their analysis of \hd, providing chemical abundances for 32 ions based on equivalent width fitting while ignoring magnetic field effects. Although these abundances are likely overestimated due to the neglect of magnetic intensification, they serve as a useful starting point for analysing this star. Using this abundance table, we computed a grid of \textsc{LLmodels} \citep{shulyak:2004} atmospheres that account for the non-solar chemical composition of the stellar surface layers. The resulting theoretical spectral energy distributions were then compared with the \textit{Gaia} DR3 externally calibrated BP/RP spectrophotometry of \hd\ \citep{montegriffo:2023}. By adopting $E(B-V) = 0.023 \pm 0.022$ \citep{lallement:2022}, fixing \logg\,=\,4.0, and using the \textit{Gaia} DR3 parallax $\pi = 2.412 \pm 0.036$~mas, we confirmed \teff\,=\,$13\,500\pm500$~K and derived a stellar radius of $R=2.18\pm0.14~R_\odot$.

To refine the stellar rotational period, we analysed five sectors of \textit{TESS} \citep{ricker:2015} photometric observations of \hd, spanning from January 2019 (sector 7) to December 2024 (sector 87). The light curves were extracted from the \textit{TESS} full-frame images using the \textsc{TESScut} code \citep{brasseur:2019} and analysed with the custom time-series analysis tools described in \citet{kochukhov:2021b}. This yielded a rotational period of $P_{\rm rot} = 7.8889898(38)$~d, approximately 14~s shorter than the value derived by \citetalias{hubrig:2024} from a subset of \textit{TESS} data. This discrepancy is negligible for the relative phasing of the HARPSpol spectra. Consequently, for the Stokes profile modelling presented below, we adopted the same rotational ephemeris as \citetalias{hubrig:2024}.

The projected rotational velocity, \vsini\,=\,$13.8\pm0.5$~\kms, was estimated by fitting theoretical spectra computed with the \textsc{Synmast} code \citep{kochukhov:2010a} to the observed profiles of the \ion{Fe}{ii} 449.140~nm line, which has a low Land\'e factor ($g_{\rm eff} = 0.43$). By combining information on $P_{\rm rot}$, $R$, and \vsini, we determined the stellar inclination using the Bayesian formalism of \citet{bowler:2023}. This analysis yielded $i = 77.6{\substack{+8.4\degr \\ -9.6\degr}}$, and we adopted $i = 75\degr$ for the ZDI inversions presented below.

\section{Multiline polarisation analysis}

The magnetic field of \hd\ is strong compared to typical Ap/Bp stars, producing clear Stokes $V$ signatures in many individual spectral lines. However, the available HARPSpol observations of this star are characterised by a relatively low signal-to-noise ratio (S/N) of 90--120. To enhance the data quality and isolate spectropolarimetric signatures of individual species, we applied the least-squares deconvolution \citep[LSD,][]{kochukhov:2010a} technique, allowing us to obtain high-S/N mean Stokes $I$ and $V$ profiles for several chemical elements.

Using the stellar parameters discussed in Sect.~\ref{sec:params} and the abundances from \citetalias{hubrig:2024}, we extracted a tailored atomic line list for \hd\ from the VALD database \citep{ryabchikova:2015}. We then applied the multiprofile version of LSD, as described by \citet{kochukhov:2010a}, to different groups of spectral lines. The best results were obtained for the line masks of Cr, Ti, and Fe, which included 199, 251, and 2268 spectral lines, respectively. The resulting LSD profiles, shown in Fig.~\ref{fig:zdi_prf}, were computed with a normalisation wavelength of $\lambda=500$~nm, an effective Land\'e factor of $g_{\rm eff}=1.2$, and a line depth of $d=0.4$.

\begin{figure}
\centering
\figps{\hsize}{0}{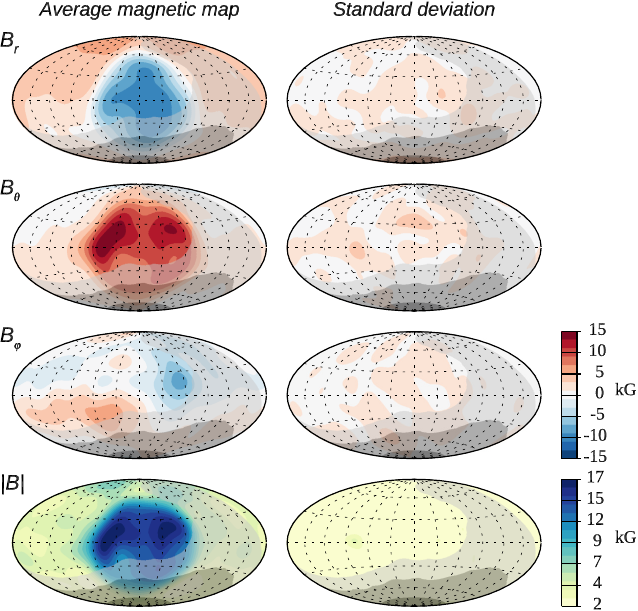}
\caption{Results of the ZDI analysis of \hd. The right column shows the magnetic field topology derived by averaging ZDI maps inferred from the Fe, Cr, and Ti LSD profiles. The left column presents the corresponding standard deviation maps. The four rows display Hammer-Aitoff projections of the radial, meridional, and azimuthal components of the magnetic field, as well as the field modulus. The greyscale overlay represents the relative visibility of different surface regions, considering stellar inclination and rotational phase coverage \citep{kochukhov:2022}. The darkest region is entirely invisible, while lighter shades correspond to relative visibilities of $\le$\,25 and $\le$\,50 per cent.}
\label{fig:zdi_avg}
\end{figure}

\section{Zeeman Doppler imaging}
\label{sec:zdi}

The ZDI analysis of \hd\ was performed using the code and methodology introduced by \citet{kochukhov:2014}, which was later refined and applied to various early-type stars in \citet{kochukhov:2017a,kochukhov:2019,kochukhov:2022,kochukhov:2023a}. In summary, we conducted detailed polarised radiative transfer calculations to generate local spectra in all four Stokes parameters across the entire HARPSpol wavelength range, considering different field strengths, field vector orientations, limb angles, and element abundances. The LSD procedure was then applied to these calculations using the same wavelength-dependent weights and line masks as in the observational analysis. The resulting grid of local Stokes $I$ and $V$ LSD profiles was used to generate disk-integrated LSD profiles for the prescribed magnetic and chemical maps. This forward model was incorporated into the \textsc{InversLSD} code, allowing us to derive the magnetic field distribution and chemical abundance maps without the usual simplifying assumptions inherent in LSD-based ZDI analyses \citep[e.g.][]{morin:2008,folsom:2018}. The inversions were carried out using Ti, Cr, and Fe LSD profiles separately, yielding three independent surface field maps. In each case, the magnetic field was represented using a general spherical harmonic expansion \citep{kochukhov:2014}, extended to $\ell_{\rm max}=10$.

The average magnetic field map of \hd\ is presented in Fig.~\ref{fig:zdi_avg}, alongside the standard deviation maps. The corresponding individual magnetic field maps are shown in Fig.~\ref{fig:zdi_all}, while the ZDI fits to the observed Stokes $I$ and $V$ profiles of the three elements are illustrated in Fig.~\ref{fig:zdi_prf}. This analysis consistently reveals a simple bipolar magnetic field distribution for \hd, as indicated by the radial field polarity. However, the local field strength deviates significantly from a dipolar geometry. The negative (inward-directed) magnetic field region reaches up to 17.8~kG, whereas the positive field region does not exceed 11.6~kG. This field strength asymmetry is counterbalanced by a larger fraction (66 per cent) of the stellar surface being occupied by the positive field. Notably, the strong-field region associated with the negative radial field is dominated by horizontal, predominantly meridional, magnetic fields.

\begin{figure}
\centering
\figps{0.8\hsize}{0}{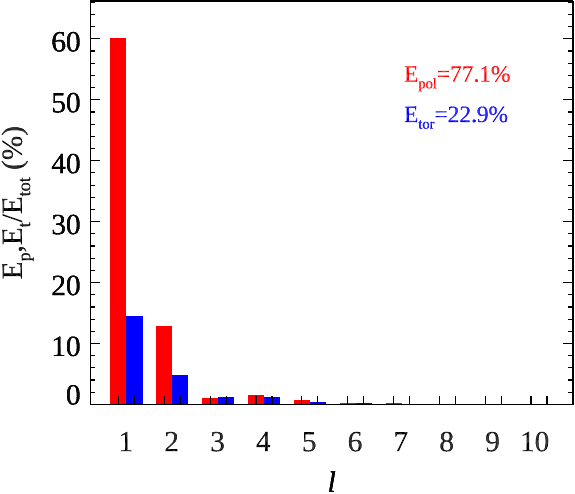}
\caption{Fractional energies of the poloidal (red) and toroidal (blue) harmonic components as a function of the angular degree $\ell$.}
\label{fig:harm}
\end{figure}

As illustrated by the magnetic energy spectrum in Fig.~\ref{fig:harm}, despite its topological peculiarities, the stellar field geometry remains largely dipolar ($\ell=1$ modes containing 75 per cent of the magnetic energy) and poloidal (77 per cent of the energy concentrated in poloidal modes). Thus, the surface field of \hd\ represents another example of a distorted dipolar configuration, though with more pronounced asymmetries than those typically observed in ZDI studies of early-type stars. \cla{However, the magnetic field of \hd\ clearly does not exhibit the same level of complexity as those of HD\,37776 or $\tau$~Sco, which are the only known early-type stars with distinctly non-dipolar magnetic field topologies.}

The standard deviation maps indicate average uncertainties of 0.7--3.6~kG for the three magnetic field vector components and the field modulus. This corresponds to 6--13 per cent of the peak field values in the respective maps. Abundance distributions (not shown here) reconstructed alongside the magnetic field maps reveal relatively little surface variation, with maximum abundance contrasts of just 0.7--1.1~dex compared to 2--4~dex often observed for Ap/Bp stars. This suggests that the prominent Stokes $I$ LSD profile variability observed in Fig.~\ref{fig:zdi_prf} is primarily driven by the magnetic field.

\section{Discussion}

According to our ZDI results, the global surface-averaged magnetic field strength of \hd\ is 7.6~kG, with the local field strength ranging from 2.1 to 17.8~kG. The average field geometry model predicts rotational variation of the mean field modulus between 4.4 and 12.8~kG. This is lower than the maximum \bs\ of 15.6~kG reported by \citetalias{hubrig:2024}. However, their Gaussian-based measurements of Zeeman-split lines may be inaccurate for such a strong-field star.

\begin{figure}
\centering
\figps{0.9\hsize}{0}{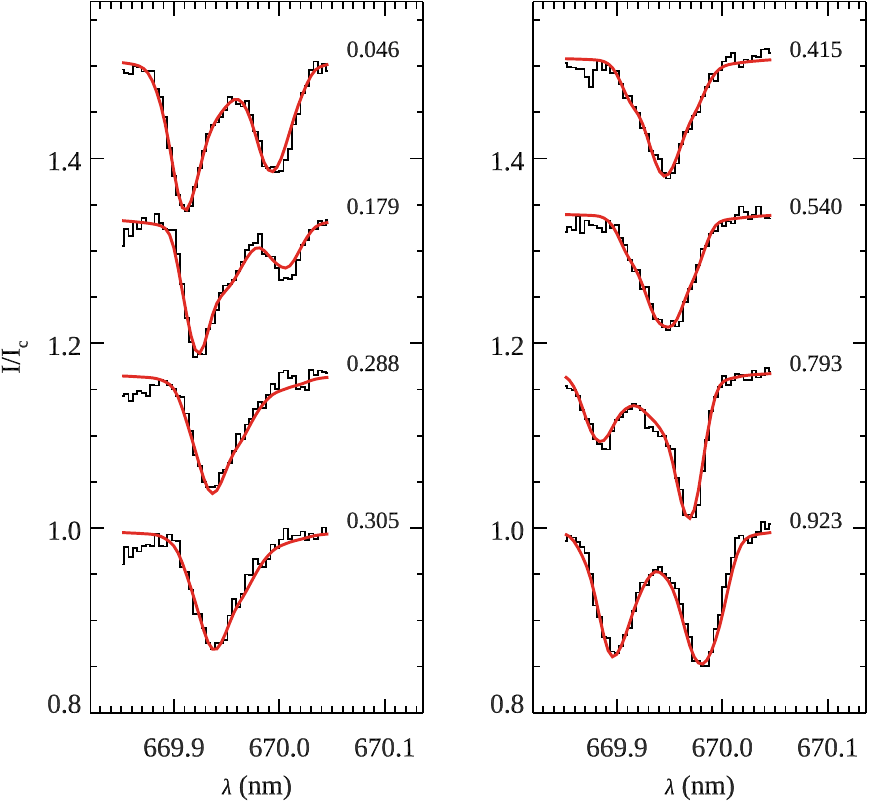}
\caption{Observed (histograms) and best-fitting ZDI model (solid lines) intensity profiles of the \ion{Si}{ii} 669.943~nm spectral line. Observations corresponding to different nights are offset vertically, with the rotational phases indicated to the right of each spectrum.}
\label{fig:si_line}
\end{figure}

As an alternative approach to validating the ZDI magnetic field map, we calculated the Stokes $I$ spectra around the \ion{Si}{ii} 669.943~nm line, which exhibits resolved Zeeman components in half of the HARPSpol observations. We employed the \textsc{Invers10} code \citep{piskunov:2002a} to adjust the Si abundance while keeping the magnetic field map fixed according to the results from Sect.~\ref{sec:zdi}. The resulting excellent fit to the observed profiles of this line, shown in Fig.~\ref{fig:si_line}, confirms that the ZDI model is consistent with the profiles of Zeeman-split lines in the optical.

Extension of this analysis to the H-band, modelling some of the lines with resolved Zeeman components identified in this region by \citetalias{hubrig:2024}, would be particularly interesting. However, such an analysis requires near-infrared observational data of higher quality than the medium-resolution APOGEE spectra with a sparse phase coverage currently available for this star.

Considering our results in the broader context of magnetism studies of early-type stars, the field structure of \hd\ represents an intriguing example of a rare, strongly asymmetric bipolar configuration. This structure exhibits an extreme contrast between the two poles while retaining the fundamental field orientation pattern of a dipolar geometry. Thus, stars with a large ratio of maximum to minimum \bs, similar to \hd, may harbour such asymmetric dipoles rather than field geometries with significant quadrupolar or higher-order contributions.

Comparing the findings of this study with previous ZDI analyses of Ap/Bp stars, we conclude that the field structure of \hd\ is not entirely unique but shares some similarities with the magnetic field topology of the Bp star HD\,133880 (HR\,5624), studied by \citet{kochukhov:2017a}. The magnetic field geometries of these two stars are compared side by side, along with a perpendicular dipolar field, in Fig.~\ref{fig:comp}. Both stars exhibit a compact, strong negative field region and a larger, less well-defined positive field that is not associated with any extremum in the field modulus map. Notably, these similar field topologies emerge from two distinct ZDI studies. Unlike \hd, HD\,133880 rotates very rapidly, providing exceptional Doppler resolution of its surface, and was studied with an extensive data set comprising over 50 Stokes $V$ spectra. The consistency of these ZDI results obtained under different conditions argues against interpreting the asymmetric dipolar geometry as an inversion artefact. Instead, this structure appears to be a genuine feature of the magnetic topology in some late-B magnetic stars and is likely more common than, for example, quadrupole-dominated fields, for which no confirmed examples currently exist. We emphasise the importance of considering this asymmetric bipolar topologies in empirical modelling of magnetic observables, ranging from integral quantitates, such as \bz\ and \bs, to Stokes parameter profiles. At the same time, theoretical studies of magnetism in early-type stars should also address the existence and implications of these magnetic geometries.

\begin{acknowledgements}
The author acknowledges support by the Swedish Research Council (grant agreement no. 2023-03667) and the Swedish National Space Agency.
\end{acknowledgements}

%\bibliographystyle{aa}
%\bibliography{astro_papers}

\appendix

\onecolumn 

\section{ZDI results for individual chemical elements}

\begin{figure*}[!h]
\centering
\figps{0.7\hsize}{0}{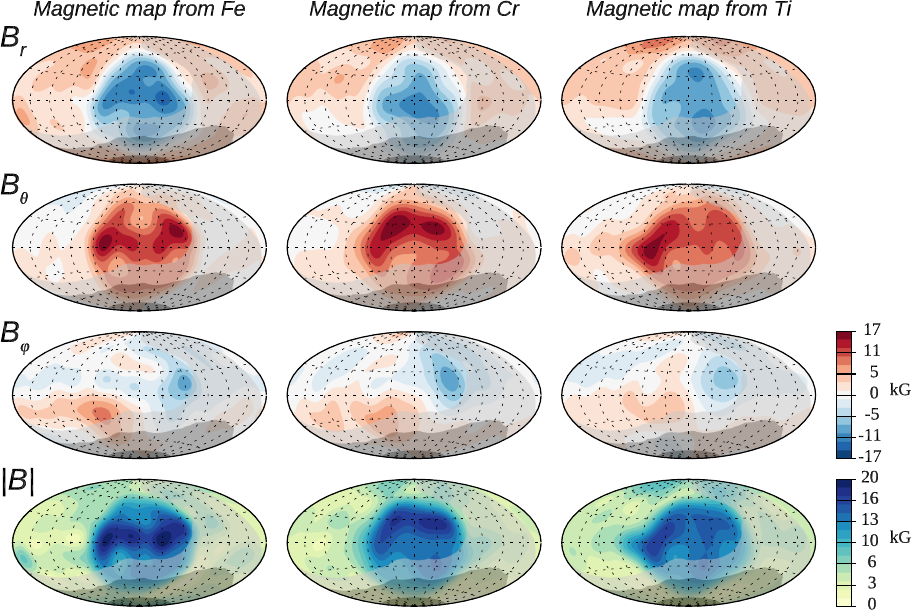}
\caption{Same as Fig.~\ref{fig:zdi_avg} but for the individual magnetic field maps of \hd\ reconstructed from the Fe, Cr, and Ti LSD profiles.}
\label{fig:zdi_all}
\end{figure*}

\begin{figure*}[!h]
\centering
\figps{0.90\hsize}{0}{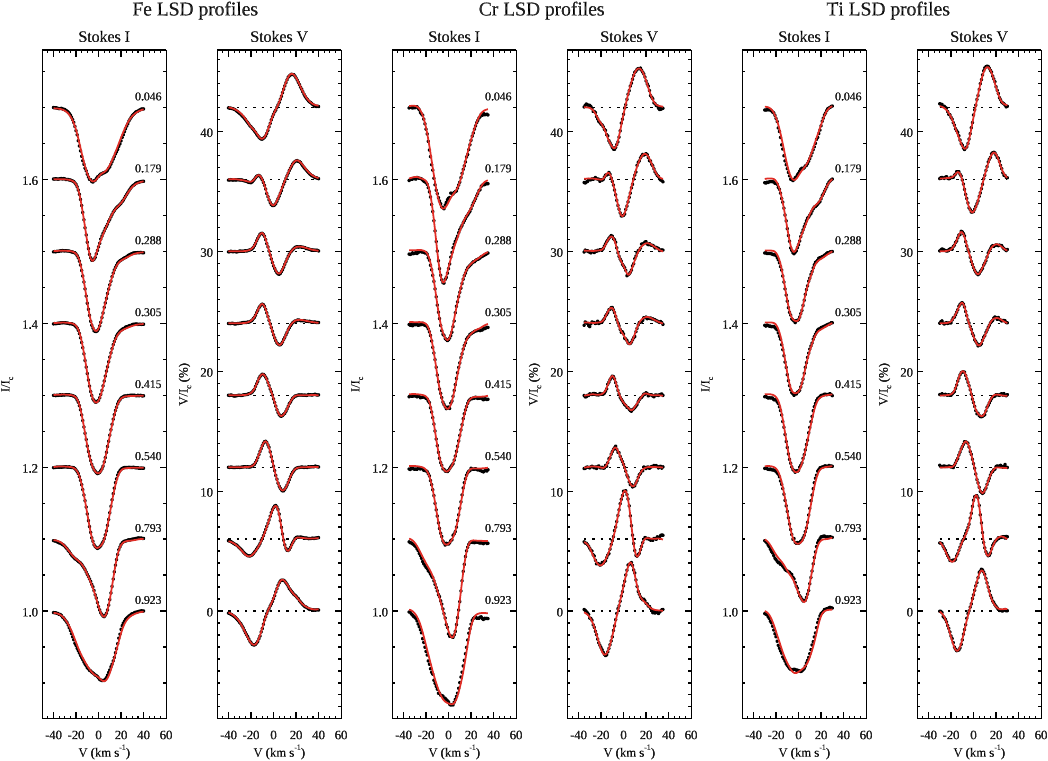}
\caption{Comparison of the observed (symbols) and best-fitting ZDI model (solid lines) LSD Stokes $I$ and $V$ profiles of \hd. The figure shows three pairs of profiles for Fe, Cr, and Ti respectively. For each pair, the left panel corresponds to Stokes $I$ and the right one to Stokes $V$. The data for different observing nights are offset vertically, with the rotational phases indicated next to each Stokes $I$ profile.}
\label{fig:zdi_prf}
\end{figure*}

\section{Comparison of \hd\ and HD\,133880}

\begin{figure*}[!h]
\centering
\figps{0.31\hsize}{0}{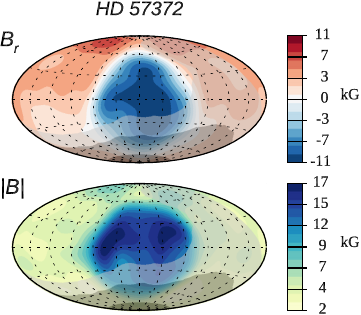}\hspace*{5mm}
\figps{0.31\hsize}{0}{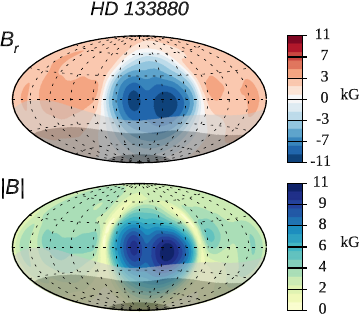}\hspace*{5mm}
\figps{0.31\hsize}{0}{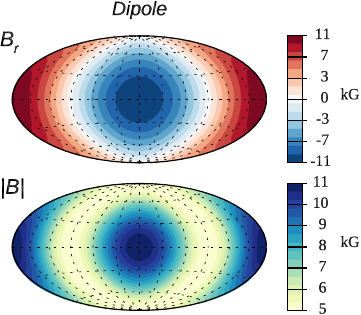}
\caption{Same as Fig.~\ref{fig:zdi_avg} but for the radial field component and the field modulus of \hd\ (left column) compared to the ZDI results for HD\,133880 (middle column) from \citet{kochukhov:2017a}. For reference, the right column shows magnetic maps for a centred dipolar field with a polar strength of $B_{\rm d}$\,=\,11~kG and an obliquity of $\beta$\,=\,90\degr.}
\label{fig:comp}
\end{figure*}

\end{document}